\documentclass[11pt,dvips,twocolumn]{article}

\usepackage{epsfig}
\usepackage{picinpar}
\usepackage{floatflt}
\makeatletter
\renewcommand{\section}{\@startsection{section}{1}{0in}
	{0.4\baselineskip}{0.1\baselineskip}{\Large\bf}}
\renewcommand{\subsection}{\@startsection{subsection}{2}{0in}
	{0.25\baselineskip}{-\baselineskip}{\large\bf}}
\renewcommand{\subsubsection}{\@startsection{subsubsection}{3}{0in}
	{0.1\baselineskip}{-\baselineskip}{\normalsize\bf}}
\makeatother
\begin{document}
\onecolumn
%
%
\makeatletter\newcommand{\ps@icrc}{
\renewcommand{\@oddhead}{\slshape{HE 2.2.16}\hfil}}
\makeatother\thispagestyle{icrc}

\begin{center}
%
{\LARGE \bf Correlations between parameters of extended air showers and their proper use in analyses}
\end{center}

\begin{center}
%
%
{\bf Wolfgang Wittek, Harald Kornmayer}\\
{\it Max-Planck-Institut f\"ur Physik, M\"unchen}\\
for the HEGRA Collaboration
\end{center}

\begin{center}
{\large \bf Abstract\\}
\end{center}
\vspace{-0.5ex}
%
%
%

\vspace{1ex}

In air shower experiments information about the initial cosmic ray particle or about the shower development is obtained by exploiting the correlations between the quantities of interest and the directly measurable quantities. It is shown how these correlations are properly treated in order to obtain unbiased results.  As an example, the measurement of the average penetration depth as a function of the shower energy is presented.

\section{Introduction}

In air shower experiments cosmic ray particles are studied in an indirect way: the distributions of the interesting quantities ($\rm \vec{X}_{orig}$) of the initial cosmic ray particle (like its nature and energy) or of the air shower (like the penetration depth) have to be inferred from the distributions of measurable quantities ($\rm \vec{X}_{meas}$) (like particle and $\check{\rm C}$-light densities at detector level).
Using Monte Carlo (MC) simulations, in which the interaction of the cosmic ray particle with the atmosphere, the shower development and also the properties of the detector are simulated, one is able to establish the correlations between the measurable quantities $\rm \vec{X}_{meas}$ and the interesting quantities $\rm \vec{X}_{orig}$.
These correlations are then used to determine the distributions of $\rm \vec{X}_{orig}$ from the experimental distributions of $\rm \vec{X}_{meas}$.

 The aim of this paper is to demonstrate that it is essential to treat the correlations in a mathematically correct way in order to avoid biases in the results. As an example the determination of the average penetration depth $\rm X_{max}$ of air showers in the atmosphere as a function of the shower energy E is presented, using data from the HEGRA array of scintillator and wide-angle $\check{\rm C}$-light detectors. 

In this example, the number of shower particles $({\rm N}_{\rm s})$ and the $\check{\rm C}$-light radius at detector level $({\rm R}_{\rm L})$ are taken as measurable quantities. ${\rm N}_{\rm s}$ is determined from the particle densities as measured by the matrix of scintillation detectors and ${\rm R}_{\rm L}$ is obtained as the inverse of the slope of the lateral $\check{\rm C}$-light distribution as measured by the matrix of $\check{\rm C}$-light detectors. The quantities of interest (''true'' quantities) are the shower energy E and the penetration depth $\rm X_{max}$.

In first approximation, ${\rm N}_{\rm s}$ carries mainly information on E, and ${\rm R}_{\rm L}$ mainly on $\rm X_{max}$. While the correlation between ${\rm R}_{\rm L}$ and $\rm X_{max}$ is quite independent of the nature (or atomic number A) of the cosmic ray particle (Lindner, 1998), this is not the case for the correlation between ${\rm N}_{\rm s}$ and E. It has been shown in (Cortina, 1997) that by using a modified ${\rm N}_{\rm s}$ ($\rm N_{s}^{c} = N_s \cdot {R_L}^\alpha$, with $\alpha$ depending on the zenith angle) the correlation between $\rm N_{s}^{c}$ and E is quite independent of A.

\section{The Method}

The task is now to determine the 2-dimensional distribution of the variables (log E, ${\rm X_{max}}$) from the 2-dimensional distribution of the variables (log $\rm N_{s}^{c}, 1/R_L$). The simplest way of doing this is to use the average correlations
\par\vspace{-0.3cm}
$$\rm < log\; N_{s}^{c} >  =  f_1 (log\; E );\qquad
 < 1/R_L > = f_2 (X_{max})$$
as determined from a sample of Monte Carlo events. In this procedure a one-to-one correlation is assumed between log E and log $\rm N_{s}^{c}$ and between $\rm X_{max}$ and $\rm 1/R_L $ respectively. In addition, a possible $\rm X_{max}$ dependence of the log E $\rm -\; log\; N_{s}^{c}$ correlation, and a log E dependence of the
 $\rm X_{max} - 1/R_L $
\newline correlation is ignored. This procedure will later be referred to as the ''one-to-one correlation procedure''.

A mathematically correct approach is to use the full correlation between (log E, $\rm X_{max}$) and\\ (log $\rm N_{s}^{c}, 1/R_L$). It is convenient to define a grid in the (log E, $\rm X_{max}$) plane and a grid in the (log $\rm N_{s}^{c},\\ 1/R_L$) plane, and some numberings $\rm (i = 1\; to\; N_{xy})$ and $\rm (j = 1\; to\; N_{uv})$ of the resulting bins in the two planes respectively.  The bin content of bin (j, i) of the distribution (log $\rm N_{s}^{c}, 1/R_L$) versus (log E, $\rm X_{max}$) for the sample of Monte Carlo events may be denoted by $\rm G _{ji}$. The full correlation can then be written as
\par\vspace{-0.4cm}
\begin{equation} 
\label{eqa}
\mathrm {g_{ji} = \frac{G_{ji}}{(\sum_k\;G_{ki})}}
\end{equation}
$\rm g _{ji_{o}}$ describes how a particular pair of values (log E, $\rm X_{max}$), defined by a specific value $\rm i_{o}$ of i, is transformed into a distribution of (log $\rm N_{s}^{c}, 1/R_L$), given by $\rm g _{ji_{o}}$ (j = 1 to $\rm N_{uv}$). The division by ($\sum_{\rm k} G_{\rm ki}$) was done to make $\rm g _{ji}$ independent of the (log E, $\rm X_{max}$) distribution in the Monte Carlo sample.

If the experimental distribution of 
(log $\rm N_{s}^{c}, 1/{R_{L}}$) is denoted by ${\rm a_{j}\; ({\rm j = 1\; to\; N_{uv}}}$) and the distribution to be determined in (log E, $\rm X_{max}$) by $\rm b _{i}$ $\rm (i = 1\; to\; N_{xy})$, then $\rm b_{i}$ has to fulfill the condition
\par\vspace{-0.26cm}
\begin{equation}
\label{eqb}
\mathrm{a_{j} = \sum_i(g_{ji}\cdot b_{i})}
\end{equation}
\par\vspace{-0.26cm}
\noindent
The condition (\ref{eqb}) ensures that the full correlations between the measured and the true quantities are taken into account.

Determining the distribution $\rm b _{i}$ from a measured distribution 
$\mathrm{ a_{j}}$, with known response matrix $\rm g_{ji}$, is a typical unfolding problem. The main point of the unfolding methods is to impose, in addition 
to (\ref{eqb}), certain smoothness conditions on the distribution $\rm b_{i}$ in order to avoid strong fluctuations of $\rm b_{i}$, which arise from statistical fluctuations of $\rm a_{j}$. In the example discussed here the method of reduced crossed entropy (MRX) is applied (Schmelling, 1994). In this method, a kind of smoothness condition is imposed by requiring the solution $\rm b_{i}$ to be close to a prior distribution $\rm b_{i}^{prior}$. $\rm b_{i}^{prior}$ may be some guess of the true distribution. Usually the result $\rm b_{i}$ is quite independent of the choice of $\rm b_{i}^{prior}$, so that $\rm b_{i}^{prior}$ may be set to a constant.

\section{Results}

The response matrix $\rm g_{ji}$ for the example discussed here is shown in Fig. 1. The ordinate corresponds to bins in the (log E, $\rm X_{max}$) plane, the abscissa to bins in the (log $\rm N_{s}^{c}, 1/{R_{L}}$) plane. 
$\rm g_{ji}$ was obtained by averaging the response matrices for different chemical elements (A = $\rm 1,\;4,\;16\; and\;56$) and by smoothing the average response matrix in the following way: $\rm g_{ji}$ was parametrized as a 2-dimensional Gaussian distribution in the variables (u, v) = (log $\rm N_{s}^{c}, 1/{R_{L}}$)
\begin{equation}
\label{eqc}
\mathrm{ g_{ji} = {\frac{1}{2\pi\sigma_{u}\sigma_{v}\;\sqrt{1 - \rho^{2}}}}\;\cdot\;
exp\Big\{ - {\frac{1}{2(1-\rho^2)}} \Big[ \Big( {\frac{u - \bar{u}}{\sigma_u}}\Big) ^2\;
                      -\;2\rho {\frac{(u - \bar{u}) (v - \bar{v})}{\sigma_u \sigma_v}}\; 
                 +\;\Big( {\frac{v - \bar{v}}{\sigma_v}}\Big) ^2\Big] \Big\} }
\end{equation}
where the 5 parameters p = $\rm \bar{u},\bar{v}, \sigma _{u}, \sigma _{v}$ and $\rho$ where assumed to be linear functions of log E and $\rm X_{max}$ (3 parameters for each of the 5 parameters p). The $\rm 5\; x\; 3 = 15$ free parameters were determined by fitting expression (\ref{eqc}) to the average response matrix. The fitted values of the parameters characterize in detail the behaviour of the correlations between (log E, $\rm  X_{max}$) and (log $\rm N_{s}^{c}, 1/{R_{L}}$) and their dependence on log E and $\rm X_{max}$. In particular one finds:
In very good approximation, $\rm < 1/R_{L} > (= \bar{v})$ is only a function of $\rm X_{max}$. $\rm  < log\; N_{s}^{c} > (= \bar{u})$ is mainly a function of log E, with some additional dependence on $\rm X_{max}$. The parameter $\rho$, which describes the correlation between log $\rm N_{s}^{c}\; and\;  1/{R_{L}}$ at fixed (log E, $\rm X_{max}$), is a function of log E: at low log E log $\rm N_{s}^{c}\; and\; 1/{R_{L}}$ are anti-correlated, whereas they are positively correlated at higher log E. All these properties of the correlations are, of course, taken into account in the unfolding procedure. 

The experimental distribution of (log $\rm N_{s}^{c}, 1/{R_{L}}$) is shown in Fig. 2 (Kornmayer, 1999). It should be noted that the measurements presented in this figure are based on preliminary data and are not the final official HEGRA results. By applying the MRX method a distribution of (log E, $\rm X_{max}$) is obtained (''unfolded'' distribution) which is displayed in Fig. 3. Forming the average $\rm X_{max}$ for each bin of log E yields the result for the elongation plot $\rm < X_{max} > $ versus log E, shown in Fig. 4a. In Fig. 4b the RMS of $\rm X_{max}$ is plotted as a function of log E.

For comparison, in Fig. 4 the results from the ''one-to-one correlation procedure'' (see above) are also plotted. It can be seen that the latter procedure underestimates $\rm < X_{max} > $ by $\rm \sim 30\; g / cm^{2}$ at low log E and by $\rm \sim 10\; g / cm^{2}$ at high log E. The points in the bin of highest energy should be taken with care because they are based on low statistics both in the experimental data and in the MC sample. No systematic differences between the two methods are seen for the RMS of $\rm X_{max}$ (Fig. 4b). Knowing that the one-to-one correlation procedure yields biased results one can try to correct the results by applying additional correction factors to $\rm < X_{max} > $, which are determined from MC events. However, these correction factors will in general depend on the details of the MC simulation, in particular on the distribution of (log E, $\rm X_{max}$).

By construction, the result of the unfolding procedure does not depend on the underlying MC distribution of (log E, $\rm X_{max}$). By fulfilling the condition (\ref{eqb}) (at least approximately, see Fig. 2), it takes into account the full correlations between the measured and the true quantities. Of course, these correlations and thus also the result for the elongation plot will depend on the model used in the MC simulation. How they depend on the MC model can be studied by doing the unfolding for different response matrices, corresponding to different MC models.

A study of the dependence of the results from the one-to-one correlation procedure on the MC model will be less conclusive because effects due to differences between the MC models and effects due to using a mathematical incorrect procedure are not well separated.

Since the average penetration depth of air showers depends on the
nuclear mass number A of the cosmic ray particle inducing the air shower,
a measurement of $\rm < X_{max} > $ as a function of 
E can be used to obtain
information on the chemical composition of cosmic rays (see for example
Roehring, 1999).

It should however be noted that from the 
same experimental data information about the chemical composition can also be obtained in a more direct way: one possibility is to start from the experimental 2-dimensional distribution of (log $\rm N_{s}, 1/{R_{L}}$) and apply the unfolding procedure to obtain the 2-dimensional distribution of (log E, log A). In this case the response matrix would explicitly depend on log E and log A and one would not have to rely on an A-independence of the response matrix, as was the case for the example discussed in this paper. The A-independence was necessary in order to determine the distribution of the penetration depth. If one is only interested in the chemical composition a knowledge of this distribution is not required and the 2-dimensional distribution of (log E, log A), which contains all the information about the chemical composition as a function of E, can be obtained directly.

\section*{Acknowledgements}

Fruitful discussions with M. Schmelling are gratefully acknowledged.

\vspace{1ex}
\begin{center}
{\Large\bf References}
\end{center}
\par\vspace{-0.3cm}
\noindent Lindner, A., 1998, Astrop. Phys. 8, 235\\
 Cortina, J., et al., 1997, J. Phys. G: Nucl. Part. Phys. 23, 1733\\
 Schmelling, M., 1994, Nucl. Instr. Meth. A 340, 400\\ 
 Kornmayer, H., 1999, PhD Thesis, Technische Universit\"at M\"unchen\\
 Roehring, A., et al., 1999, OG.1.2.09, 
 Proc. \rm 26th ICRC (Salt Lake City, 1999)
%
%
\twocolumn
\begin{figure}
\vspace*{-1.45cm} 
\epsfig{file=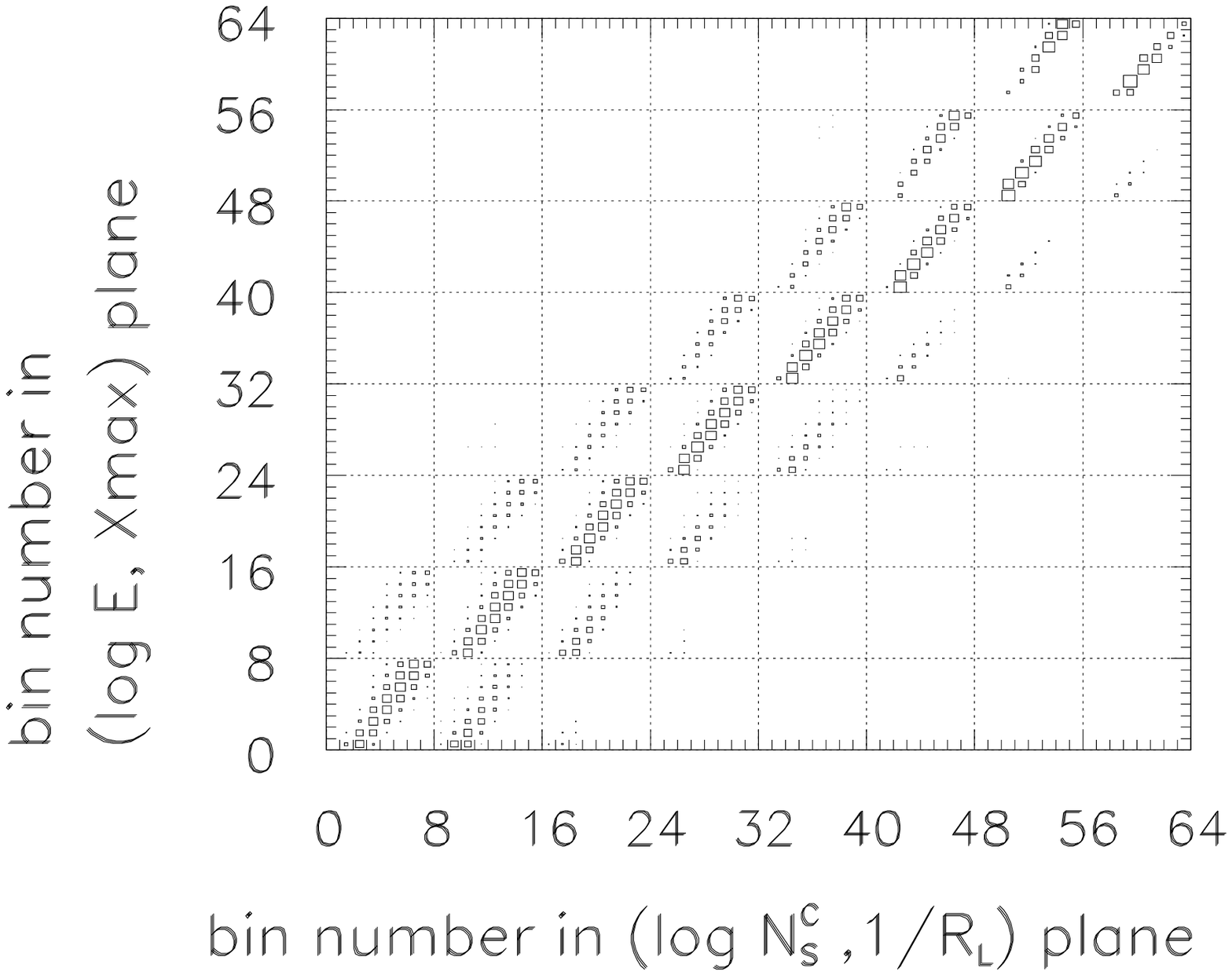,width=0.48\textwidth}
\vspace*{-0.8cm}
\caption[..]{Response matrix $\mathrm{g_{ji}}$. The abscissa corresponds to
bins in the ($\mathrm{log\; N_s^c,\; 1/R_L} $) plane: In 8 consecutive bins
$\mathrm{1/R_L} $ increases from 0.0 to 0.035 ${\rm m}^{-1}$.
Each block of  8 consecutive bins is for one bin in  $\mathrm{log\; N_s^c}$.
In 8 consecutive blocks $\mathrm{log\; N_s^c}$ increases from 7.27 to 10.07.
The ordinate corresponds
to bins in the ($\mathrm{log\; E,\; X_{max}}$) plane: In 8 consecutive bins
$\mathrm{X_{max}}$ increases from 320 to 800 $\mathrm{g/cm^2}$.
Each block of 8 consecutive bins  is for one bin in $\mathrm{log\; E}$.
In 8 consecutive blocks  $\mathrm{log\; E}$ increases from 1.5 to 4.3.}
\label{fig1}
\vspace*{-0.7cm}
\epsfig{file=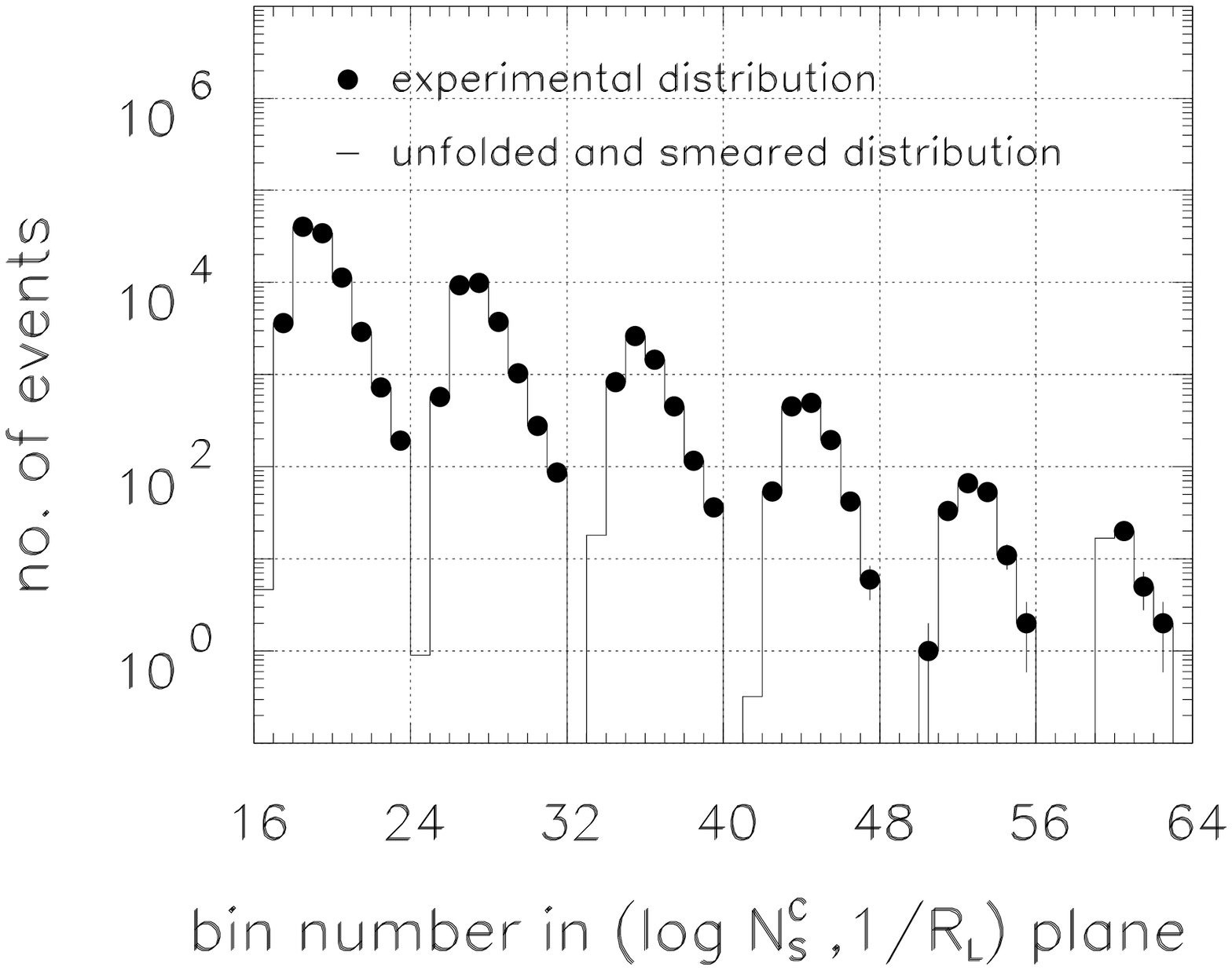,width=0.48\textwidth} 
\vspace*{-0.7cm}
\caption[..]{Experimental distribution of $\mathrm{1/R_L}$ in different
bins of $\mathrm{log\; N_s^c}$ (full circles).
The histogram represents the result from applying the response matrix 
to the unfolded distribution of ($\mathrm{log\;E,\; X_{max}}$).
The good agreement between the two distributions shows that relation (2)
between the measured and unfolded distribution is well fulfilled.
The abscissa in this figure corresponds
to the abscissa in Fig.~\ref{fig1}.}
\label{fig2} 
\end{figure}
\begin{figure}
\vspace*{-0.6cm}
\epsfig{file=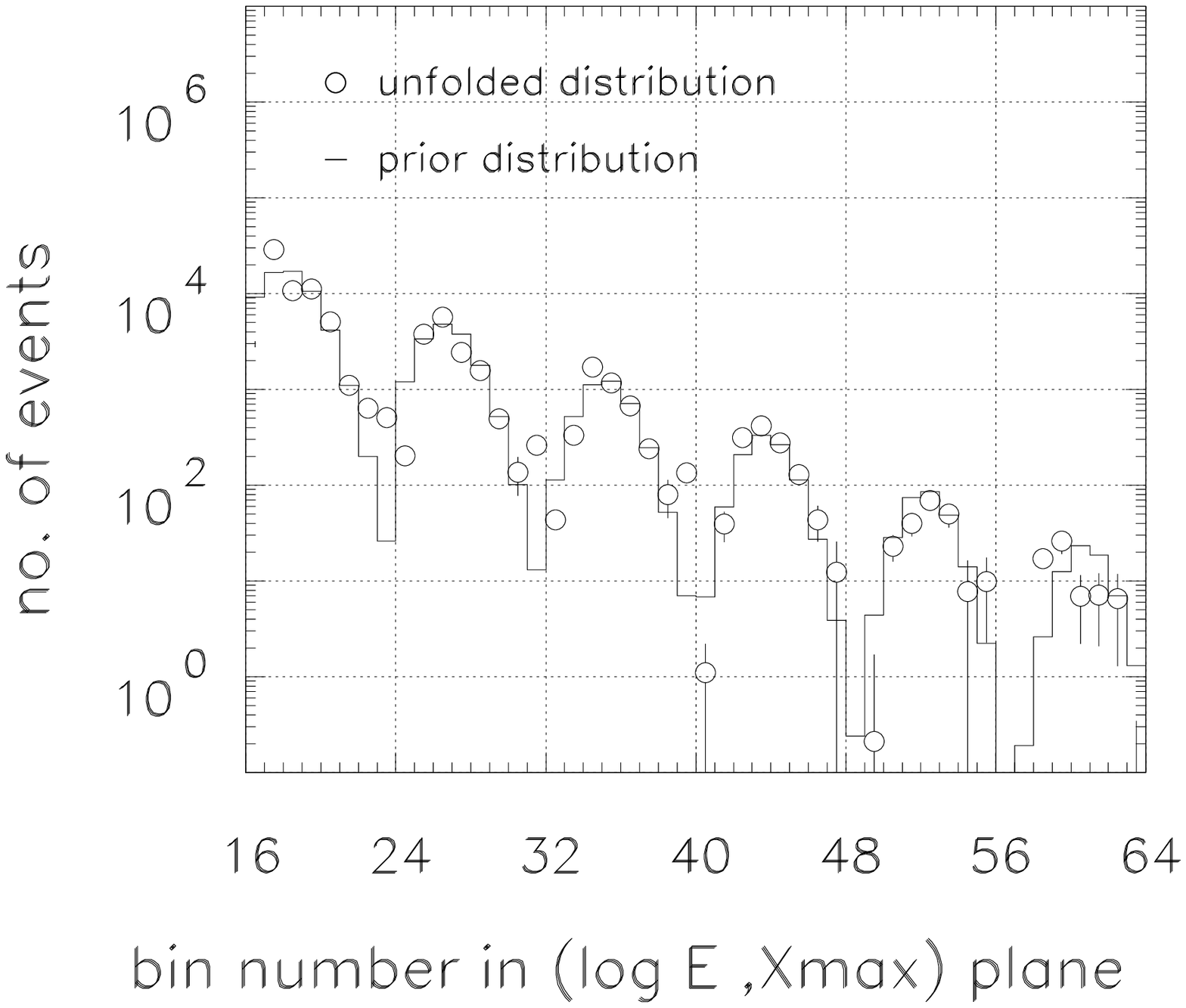,width=0.48\textwidth} 
\vspace*{-1.45cm}
 \caption[...]{Unfolded distribution
of $\mathrm{X_{max}}$ in different bins of $\mathrm{log \; E}$ (open
circles). The histogram represents the prior distribution used in the
unfolding procedure. The abscissa in
this figure corresponds to the ordinate in Fig.~\ref{fig1}.}
\label{fig3} 
\begin{center}
\vspace*{-1.7cm}
\epsfig{file=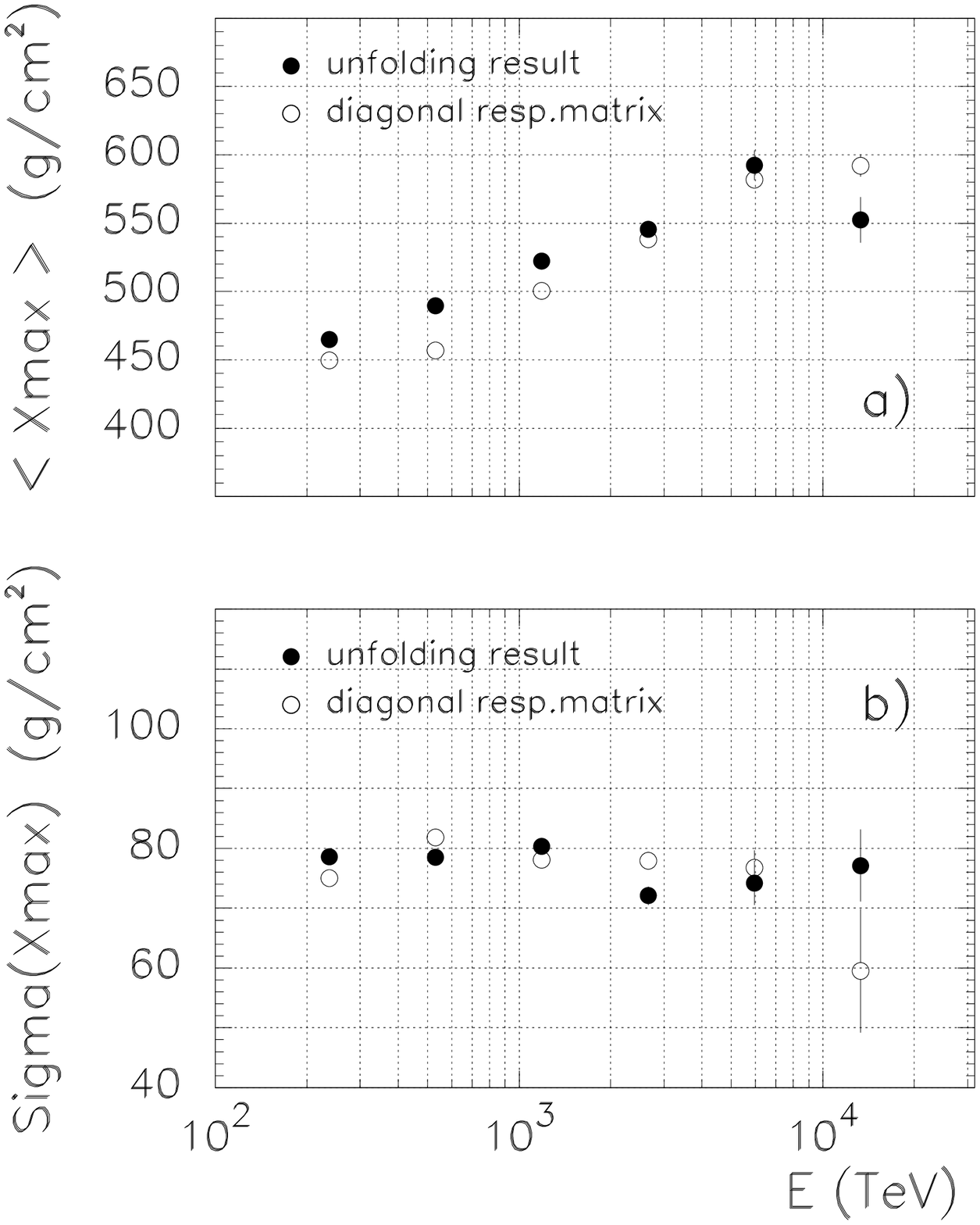,width=0.48\textwidth}
\vspace*{-0.7cm}
\caption[...]{a) Average penetration depth
$\mathrm{< X_{max}>}$ and 
b) RMS of $\mathrm{X_{max}}$ as functions  of
E. The full circles represent the results of the 
unfolding procedure using the full response matrix. If one-to-one
correlations are assumed between $\mathrm{1/R_L}$ and $\mathrm{X_{max}}$,
and $\mathrm{log\; N_s^c}$ and $\mathrm{log\; E}$ respectively, the
points represented by open circles are obtained 
(''diagonal response matrix''). While both methods yield consistent results for
RMS of $\mathrm{X_{max}}$, the one-to-one correlation procedure in general
underestimates $\mathrm{<X_{max}>}$ by 10 to \mbox{30 $\mathrm{g/cm^2}$.}}
\label{fig4}
\end{center}
\end{figure}
\end{document}